\documentclass[11pt]{article}

\newenvironment{proof}{\noindent{\bf Proof:}}{\qed\medskip}

\newtheorem{claim}{Claim}
\newtheorem{fact}[claim]{Fact}

\newtheorem{lemma}[claim]{Lemma}

\newtheorem{theorem}[claim]{Theorem}
\newtheorem{observation}[claim]{Observation}

\newcommand{\qed}{\hfill\rule{7pt}{7pt}}
\newcommand{\ol}[1]{{\tilde{#1}}}

\newcommand{\p}{{\oplus}}
\newcommand{\ra} {{\rangle}}
\newcommand{\la} {{\langle}}
\newcommand{\C}{{\cal C}}
\newcommand{\N}{{\cal N}}
\newcommand{\D}{{\cal D}}
\newcommand{\qqed}[1]{\hfill{#1}\rule{7pt}{7pt}}

\title{Quantum versus Classical Learnability}

\author{Rocco A.\ Servedio\thanks{Supported in part by an NSF graduate
fellowship and by NSF grant CCR-95-04436.}
~~~and~~~Steven J. Gortler\thanks{Supported by 
NSF Career Grant 97-03399 and the Alfred P. Sloan Foundation.
}\\
\\
Division of Engineering and Applied Sciences \\
Harvard University\\ 
Cambridge, MA 02138\\ {\tt \{rocco,sjg\}@cs.harvard.edu}}

\date{}

\begin{document}

\setcounter{page}{0}
\maketitle
\begin{abstract}

This paper studies fundamental questions in computational learning theory
from a quantum computation perspective. 
We consider quantum versions of two well-studied classical
learning models:  Angluin's model of exact learning
from membership queries and Valiant's Probably Approximately Correct (PAC) 
model of learning from random examples.  
We give positive and negative results 
for quantum versus classical learnability.
For each of the two learning models described above, we show that any concept
class is information-theoretically learnable from polynomially
many quantum examples if and only if it is information-theoretically
learnable from polynomially many classical examples.
In contrast to this information-theoretic equivalence betwen quantum
and classical learnability, though, we observe that a separation 
does exist between {\em efficient} quantum and classical learnability.
For both the model of exact learning from membership queries and the PAC 
model, we show that under a widely held computational hardness assumption
for classical computation (the intractability of factoring), 
there is a concept class which is polynomial-time
learnable in the quantum version but not in the classical version of the 
model.

\end{abstract}

\thispagestyle{empty}

\newpage

\section{Introduction}

\subsection{Motivation}
In recent years many researchers have investigated the power 
of quantum computers which can query a black-box oracle for an
unknown function \cite{BBC+98,BBB+97,BBH+98,BHT98,BrassardHoyer97,BCW98,
DeutschJozsa92,FFK+93,FortnowRogers98,Grover96,Simon97,Zalka97}.  The broad
goal of research in this area is to understand the relationship
betwen the number of quantum versus classical oracle queries which are 
required to answer various questions about the function computed
by the oracle.  For example, a well-known result due to 
Deutsch and Jozsa \cite{DeutschJozsa92} shows that exponentially
fewer queries are required in the quantum model in order
to determine with certainty whether a black-box oracle computes
a constant Boolean function or a function which is 
balanced between outputs $0$ and $1.$  More recently, 
several researchers have studied the 
number of quantum oracle queries
which are required to determine whether or not the function computed
by a black-box oracle ever assumes a nonzero value \cite{BBC+98,
BBB+97,BBH+98,BCW98,Grover96,Zalka97}.

A natural question which arises within this framework is the following:  
what is the relationship between the number of quantum versus
classical oracle queries which are required in order to {\em exactly
identify} the function computed by a black-box oracle?  Here 
the goal is not to determine whether a
black-box function satisfies some particular property 
(such as ever taking a nonzero value), but rather to precisely
identify a black-box function which belongs to some restricted
class of possible functions.  The classical version of this problem
has been well studied in the computational learning
theory literature \cite{Angluin88,BCG+96,Gavalda94,Hegedus95,HPR+96},
and is known as the problem of {\em exact learning
from membership queries}.  The question stated above can
thus be phrased as follows:  what is the relationship between the number
of quantum versus classical membership queries which are required for
exact learning?  We answer this question in this paper.

In addition to the model of exact learning from membership queries, 
we also consider a quantum version of Valiant's widely studied PAC learning
model which was introduced by 
Bshouty and Jackson \cite{BshoutyJackson99}.
While a learning algorithm in the classical PAC model has access to labeled
examples which are drawn from a fixed probability distribution,
a learning algorithm in the quantum PAC model has access to a fixed quantum
superposition of labeled examples. Bshouty and Jackson gave a polynomial-time
algorithm for a particular learning problem in the quantum PAC model,
but did not address the general relationship between the number of
quantum versus classical examples which are required for PAC learning.
We answer this question as well.

\subsection{The results}

We show that in an information-theoretic sense,
quantum and classical learning are equivalent up to polynomial
factors: for both the model of exact learning from membership queries
and the PAC model, there is no learning problem which can be solved 
using significantly fewer quantum examples than classical examples.
More precisely, our first main theorem is the following:

\begin{theorem} \label{thm:equivexact}
Let $\C$ be any concept class.  Then $\C$ is exact learnable from
a polynomial number of quantum membership queries if and only if $\C$ is 
exact learnable from a polynomial number of classical membership queries.
\end{theorem}

Our second main theorem is an analogous result for quantum
versus classical PAC learnability:

\begin{theorem} \label{thm:equivpac}
Let $\C$ be any concept class.  Then $\C$ is PAC learnable from a
polynomial number of quantum examples if and only if $\C$ is 
PAC learnable from a polynomial number of classical examples.
\end{theorem}
The proofs of Theorems \ref{thm:equivexact} and \ref{thm:equivpac} 
use several different quantum lower bound techniques and demonstrate
an interesting relationship between lower bound techniques in quantum 
computation and computational learning theory.  

Theorems \ref{thm:equivexact} and \ref{thm:equivpac} are 
information-theoretic rather than computational in nature; they show
that for any learning problem in these two models, 
if there is a quantum learning algorithm which uses 
polynomially many examples, then there must also exist
a classical learning algorithm which uses polynomially many
examples.  However, Theorems \ref{thm:equivexact} and \ref{thm:equivpac}
do not imply that every polynomial time quantum learning algorithm
must have a polynomial time classical analogue.
In fact, using known computational hardness results for classical
polynomial-time learning algorithms, we show that  
the equivalences stated in Theorems \ref{thm:equivexact}
and \ref{thm:equivpac} do {\em not} hold for efficient learnability.
Under a widely accepted computational hardness assumption
for classical computation, the hardness of factoring Blum integers,
we observe that Shor's polynomial-time factoring algorithm implies
that for each of the two learning models
considered in this paper, there is a concept class
which is polynomial-time learnable in the quantum version but
not in the classical version of the model.

\subsection{Organization}
In Section \ref{sec:prelim} we define the classical exact learning
model and the classical PAC learning model and describe the
quantum computation framework.  In Section \ref{sec:qmq} we prove
the information-theoretic equivalence of quantum and classical exact learning
from membership queries (Theorem \ref{thm:equivexact}), 
and in Section \ref{sec:qpac} we prove the information-theoretic
equivalence of quantum 
and classical PAC learning (Theorem \ref{thm:equivpac}).
Finally, in Section \ref{sec:qeff} we observe that under
a widely accepted computational hardness assumption for
classical computation, in each of these two learning models
there is a concept class which is quantum learnable in polynomial time
but not classically learnable in polynomial time.

\section{Preliminaries}
\label{sec:prelim}
A {\em concept} $c$ over $\{0,1\}^n$ is a Boolean function over the domain
$\{0,1\}^n,$ or equivalently a concept can be viewed as
a subset $\{x \in \{0,1\}^n : c(x) = 1\}$ of $\{0,1\}^n.$
A {\em concept class} $\C = \cup_{n \geq 1} C_n$ is a collection of
concepts, where $C_n  = \{c \in \C : \ c$ is a concept over 
$\{0,1\}^n\}.$  For example, $C_n$ might be the family of
all Boolean formulae over $n$ variables which are of size at most
$n^2.$ We say that a pair 
$\la x, c(x) \ra$ is a {\em labeled example} of the concept $c.$

While many different learning models have been proposed, most models 
adhere to the same basic paradigm: a learning algorithm for a concept class 
$\C$ typically has access to (some kind of) an oracle which provides  
examples that are labeled according to a fixed but unknown target concept
$c \in \C,$  and the goal of the learning algorithm is to 
infer (in some sense) the structure of the target concept $c.$  The
two learning models which we discuss in this paper, the model of exact
learning from membership queries and the PAC model, make this
rough notion precise in different ways.

\subsection{Classical Exact Learning from Membership Queries} \label{sec:mq}

The model of {\em exact learning from membership queries} was introduced by
Angluin \cite{Angluin88} and has since been widely studied
\cite{Angluin88,BCG+96,Gavalda94,Hegedus95,HPR+96}.  
In this model the learning algorithm
has access to a {\em membership oracle} $MQ_c$ where $c \in C_n$
is the unknown target concept.  When given an input string 
$x \in \{0,1\}^n,$ in one
time step the oracle $MQ_c$ returns the bit $c(x);$ such
an invocation is known as a {\em membership query} since
the oracle's answer tells whether or not $x \in c$ (viewing
$c$ as a subset of $\{0,1\}^n$). The goal 
of the learning algorithm is to construct a hypothesis
$h: \{0,1\}^n \rightarrow \{0,1\}$ which is logically equivalent to
$c,$ i.e. $h(x) = c(x) $ for all $x \in \{0,1\}^n.$  Formally, we
say that an algorithm $A$  (a probabilistic Turing machine)
is an {\em exact learning algorithm for $\C$ using membership queries}
if for all $n \geq 1,$ for all $c \in C_n,$
if $A$ is given $n$ and access to $MQ_c,$ then 
with probability at least $2/3$ algorithm
$A$ outputs a representation of a Boolean circuit $h$ such that
$h(x) = c(x)$ for all $x \in \{0,1\}^n.$ 
The {\em sample complexity} $T(n)$ of
a learning algorithm $A$ for $\C$ is the maximum number of calls
to $MQ_c$ which $A$ ever makes for any $c \in C_n.$  We
say that $\C$ is {\em exact learnable} if there is a learning 
algorithm for $\C$ which has poly$(n)$ sample 
complexity, and we say that $\C$ is {\em efficiently exact 
learnable} if there is a learning algorithm for $\C$
which runs in poly$(n)$ time.

\subsection{Classical PAC Learning} \label{sec:pac}
The PAC (Probably Approximately Correct) 
model of concept learning was introduced by Valiant in \cite{Valiant84}
and has since been extensively studied \cite{AnthonyBiggs97,KearnsVazirani94}.
In this model the learning algorithm 
has access
to an {\em example oracle} $EX(c,{\D})$ where $c \in C_n$ 
is the unknown target concept and ${\D}$ is an unknown distribution
over $\{0,1\}^n.$  The oracle $EX(c,{\D})$ takes no inputs; when
invoked, in one time step it returns a labeled example
$\la x, c(x) \ra$
where $x \in \{0,1\}^n$ is randomly selected according to the distribution
${\D}.$  The goal of the learning algorithm is
to generate a hypothesis $h: \{0,1\}^n \rightarrow \{0,1\}$ which is
an {\em $\epsilon$-approximator for $c$ under ${\D},$} i.e. a
hypothesis $h$ such that 
$\Pr_{x \in {\D}}[h(x) \neq c(x)] \leq \epsilon.$ An algorithm $A$
(again a probabilistic Turing machine)
is a {\em PAC learning algorithm for $\C$}
if the following condition holds:  for all $n \geq 1$ and $0 <
\epsilon,\delta < 1,$  for all $c \in C_n,$
for all distributions ${\D}$ over $\{0,1\}^n,$ if $A$ is given 
$n,\epsilon,\delta$ and access to $EX(c,{\D}),$ then with
probability at least $1 - \delta$ algorithm $A$ outputs a 
representation of a circuit
$h$ which is an $\epsilon$-approximator for $c$ under ${\D}.$  
The {\em sample complexity} $T(n,\epsilon,\delta)$ of a learning
algorithm $A$ for $\C$
is the maximum number of calls to $EX(c,{\D})$ which
$A$ ever makes for any concept $c \in C_n$ and any
distribution ${\D}$ over $\{0,1\}^n.$ 
We say that $\C$ is {\em PAC learnable}
if there is a PAC learning algorithm for $\C$ which has
poly$(n,{\frac 1 \epsilon},{\frac 1 \delta})$ 
sample complexity, and we say that $\C$
is {\em efficiently PAC learnable} if there is a PAC learning
algorithm for $\C$ which runs in poly$(n,{\frac 1 \epsilon},{\frac 1 \delta})$
time.

\subsection{Quantum Computation} \label{sec:qc}
Detailed descriptions of the quantum computation model can be
found in \cite{BernsteinVazirani97,Cleve99,Yao93}; here we outline 
only the basics using the terminology of {\em quantum networks} as presented
in \cite{BBC+98}.  A quantum network $\N$ is a quantum circuit 
(over some standard basis augmented with one oracle gate)
which acts on an $m$-bit quantum register; the computational
basis states of this register are the $2^m$ binary
strings of length $m.$
A quantum network can be viewed 
as a sequence of unitary transformations
$$
U_0,O_1,U_1,O_2,\dots,U_{T-1},O_T,U_T,
$$
where each $U_i$ is an arbitrary unitary transformation on $m$ qubits
and each
$O_i$ is a unitary transformation which corresponds to an oracle 
call.\footnote{
Since there is only one kind of oracle gate, each $O_i$ is 
the same transformation.}
Such a network is said to have {\em query complexity} $T.$
At every stage in the execution of the network, the current state
of the register can be
represented as a superposition $\sum_{z \in \{0,1\}^m} \alpha_z |z \ra$
where the $\alpha_z$ are complex numbers which satisfy 
$\sum_{z \in \{0,1\}^m} \|\alpha_z\|^2 = 1.$  If this state is 
measured, then with probability $\|\alpha_z\|^2$ the string $z \in \{0,1\}^m$ 
is observed and the state collapses down to $|z\ra$.
After the final transformation $U_T$ takes place, a measurement is performed
on some subset of the bits in the register and the observed value 
(a classical bit string) is the output of the computation.

Several points deserve mention here.  First, since the information
which our quantum network uses for its computation comes from the oracle
calls, we may stipulate that the initial state of the 
quantum register is always $|0^m \ra.$  Second, as described above each $U_i$
can be an arbitrarily complicated unitary transformation (as long
as it does not contain any oracle calls) 
which may require a large quantum circuit to implement.  
This is of small concern to us since we are chiefly interested in
query complexity and not circuit size.  Third, as defined above
our quantum networks can make only one measurement at the very end
of the computation; this is an inessential restriction since 
any algorithm which uses intermediate measurements can be modified
to an algorithm which makes only one final measurement.
Finally,  we have not 
specified just how the oracle calls $O_i$ work; we address this point
separately in Sections \ref{sec:qexact} and \ref{sec:qex} for
each type of oracle.

If $|\phi\ra = \sum_z \alpha_z |z \ra$ and $|\psi\ra = \sum_z \beta_z |z\ra$ 
are two superpositions of basis states, then the 
{\em Euclidean distance} betweeen $|\phi\ra$ and $|\psi\ra$ is 
$||\phi\ra - |\psi\ra| = (\sum_z |\alpha_z - \beta_z|^2)^{1/2}.$
The {\em total variation distance} between two distributions 
${\D}_1$ and ${\D}_2$ is defined to be $\sum_x |{\D}_1(x) -
{\D}_2(x)|.$  
The following fact (Lemma 3.2.6 of \cite{BernsteinVazirani97}), which
relates the Euclidean distance between two superpositions and the total 
variation distance between the distributions induced by measuring
the two superpositions, will be useful:

\begin{fact}\label{fact:3.1} 
Let $|\phi\ra$ and $|\psi\ra$ be two unit-length superpositions which
represent possible states of a quantum register.  If 
the Euclidean distance $||\phi\ra - |\psi\ra|$ is at most $\epsilon,$ 
then performing the same observation on $|\phi\ra$ and $|\psi\ra$ 
induces distributions $\D_\phi$ and $\D_\psi$
which have total variation distance at most $4 \epsilon.$
\end{fact}

\section{Exact Learning from Quantum Membership Queries} \label{sec:qmq}

\subsection{Quantum Membership Queries}
\label{sec:qexact}

A {\em quantum membership oracle} $QMQ_c$ is the natural quantum
generalization of a classical membership oracle $MQ_c$: on input 
a superposition of query strings, the oracle $QMQ_c$ generates the 
corresponding superposition of example labels. 
More formally, a $QMQ_c$ gate maps the basis state
$|x,b\ra$ (where $x \in \{0,1\}^n$ and $b \in \{0,1\}$) to the state
$|x, b \p c(x)\ra.$  If $\N$ is a quantum network which has $QMQ_c$
gates as its oracle gates, then each $O_i$ is the unitary transformation
which maps $|x,b,y \ra$ (where $x \in \{0,1\}^n,$ $b \in \{0,1\}$
and $y \in \{0,1\}^{m-n-1}$) to $|x,b \p c(x), y\ra$.\footnote{Note that
each $O_i$ only affects the first $n+1$ bits of a basis
state.  This is without loss of generality since the
transformations $U_j$ can ``permute bits'' of the network.}
Our $QMQ_c$ oracle is identical to the well-studied notion
of a quantum black-box oracle for $c$
\cite{BBC+98,BBB+97,BernsteinVazirani97,BBH+98,BHT98,BrassardHoyer97,
BCW98,DeutschJozsa92,Grover96,Zalka97}.
We discuss the relationship between our work and these results in Section 
\ref{sec:qexactcompar}.

A {\em quantum exact learning algorithm for $\C$} is a family
of quantum networks $\N_1,\N_2,\dots,$ where each
network $\N_n$ has a fixed architecture independent of the target
concept $c \in C_n,$ with the following property:  for all $n \geq 1,$
for all $c \in C_n,$ if $\N_n$'s oracle gates are instantiated
as $QMQ_c$ gates, then with probability at least $2/3$ 
the network $\N_n$ outputs 
a representation of a (classical) Boolean circuit $h: \{0,1\}^n \rightarrow \{0,1\}$
such that $h(x) = c(x)$ for all $x \in \{0,1\}^n.$  The {\em
quantum sample complexity} of a quantum exact learning algorithm for $\C$
is $T(n),$ where $T(n)$ is the query complexity of $\N_n$.
We say that $\C$ is {\em exact 
learnable from quantum membership queries} 
if there is a quantum exact learning algorithm for $\C$
which has poly$(n)$ quantum sample complexity, and we say that $\C$
is {\em efficiently quantum exact learnable} if each network 
$\N_n$ is of poly$(n)$ size.

\subsection{Lower Bounds on Classical and Quantum Exact Learning}
\label{sec:lbcqel}

Two different lower bounds are known for the number of (classical) membership queries 
which are required to exact learn any concept class.  In this section we 
prove two analogous lower bounds on the number of {\em quantum} membership queries
required to exact learn any concept class.
Throughout this section 
for ease of notation we omit the subscript $n$ and write $C$ for $C_n.$

\subsubsection{A Lower Bound Based on Similarity of Concepts}

Consider a set of concepts which are all ``similar'' in the sense that
for every input almost all concepts in the set agree.  Known
results in learning theory state that such a concept class must 
require a large number of membership queries for exact learning.
More formally, let $C^\prime \subseteq C$ be any subset of $C.$
For $a \in \{0,1\}^n$ and $b \in \{0,1\}$ let $C^\prime_{\la a,b \ra}$ denote
the set of those concepts in $C^\prime$ which assign label $b$ to example
$a,$ i.e. $C^\prime_{\la a,b \ra} = \{c \in C^\prime: c(a) = b\}.$  
Let $\gamma^{C^\prime}_{\la a,b \ra} = |C^\prime_{\la a,b \ra}|/|C^\prime|$ 
be the fraction of such concepts in $C^\prime,$
and let $\gamma^{C^\prime}_{a} = \min\{\gamma^{C^\prime}_{\la a,0 \ra},
\gamma^{C^\prime}_{\la a,1 \ra}\};$
thus $\gamma^{C^\prime}_{a}$ is the minimum fraction of concepts
in $C^\prime$ which can be eliminated by querying $MQ_c$ on the string $a.$
Let $\gamma^{C^\prime} = 
\max\{\gamma^{C^\prime}_a : a \in \{0,1\}^n\}.$  
Finally, let $\hat{\gamma}^C$ be the minimum of $\gamma^{C^\prime}$
across all $C^\prime \subseteq C$ such that $|C^\prime| \geq 2.$
Thus 
$$
\hat{\gamma}^{C} = 
\min_{C^\prime \subseteq C, |C^\prime| \geq 2} \ \ 
\max_{a \in \{0,1\}^n} \ 
\min_{b \in \{0,1\}} \  
{\frac {|{C^\prime}_{\la a,b \ra}|} {|C^\prime|}}.
$$
Intuitively, the inner $\min$ corresponds
to the fact that the oracle may provide a worst-case response to 
any query; the $\max$ corresponds to the fact that the learning algorithm
gets to choose the ``best'' query point $a;$ and the outer $\min$ corresponds 
to the fact that the learner must succeed no matter what subset $C^\prime$
of $C$ the target concept is drawn from.  Thus $\hat{\gamma}^C$ is small if
there is a large set $C^\prime$ of concepts which are all very similar 
in that any query eliminates only a few concepts from $C^\prime.$
If this is the case then many membership queries should be required
to learn $C;$ formally, we have the following lemma
which is a variant of Fact 2 from \cite{BCG+96} (the proof is given
in Appendix \ref{sec:pfs}):

\begin{lemma} \label{lem:gamma}
Any (classical) exact learning algorithm for $C$ must have sample 
complexity $\Omega({\frac 1 {\hat{\gamma}^C}}).$
\end{lemma}

We now develop some tools which will enable us to prove a quantum
version of Lemma \ref{lem:gamma}.
Let $C^\prime \subseteq C, |C^\prime| \geq 2$ be such that
$\gamma^{C^\prime} = \hat{\gamma}^C.$  
Let $c_1,\dots,c_{|C^\prime|}$ be a listing of 
the concepts in $C^\prime.$
Let the {\em typical concept for $C^\prime$} be the
function $\hat{c}: \{0,1\}^n \rightarrow \{0,1\}$ defined as
follows:  for all $a \in \{0,1\}^n,$ $\hat{c}(a)$ is
the bit $b$ such that $|C^\prime_{\la a,b \ra}| \geq |C^\prime|/2$
(ties are broken arbitrarily; note that a tie occurs only
if $\hat{\gamma}^C = 1/2$).
The typical concept $\hat{c}$ need not belong to $C^\prime$ or even 
to $C.$   Let the {\em difference matrix} $D$ be the
$|C^\prime| \times 2^n$ zero/one matrix where 
rows are indexed by concepts in $C^\prime,$ 
columns are indexed by strings in $\{0,1\}^n,$ and $D_{i,x} = 1$ iff 
$c_i(x) \neq \hat{c}(x).$  By our choice of $C^\prime$ and 
the definition of $\hat{\gamma}^C,$ 
each column of $D$ has at most $|C^\prime| \cdot \hat{\gamma}^C$ ones, 
i.e. the $L_1$ matrix norm of $D$ is $\|D\|_1 \leq |C^\prime| \cdot 
\hat{\gamma}^C.$

Our quantum lower bound proof uses ideas which were first introduced by 
Bennett et al.  \cite{BBB+97}.
Let $\N$ be a fixed quantum network architecture and let 
$U_0,O_1,\dots,U_{T-1},O_T,U_T$ be the corresponding sequence
of transformations.  For $1 \leq t \leq T$ 
let $|\phi_t^c \ra$ be the state of the quantum register after the 
transformations up through $U_{t-1}$ have been performed 
(we refer to this stage of the computation as time $t$)
if the oracle gate is $QMQ_c.$
As in \cite{BBB+97}, for $x \in \{0,1\}^n$ let $q_x(|\phi_t^c \ra),$ 
the {\em query magnitude of string $x$ at time $t$ with respect to $c$,} be the sum of the 
squared magnitudes in $|\phi_t^c \ra$ of the basis states 
which are querying $QMQ_c$ on string $x$ at time $t;$ so
if $|\phi_t^c \ra = \sum_{z \in \{0,1\}^m} \alpha_z |z \ra,$ then
$$
q_x(|\phi_t^c\ra) = 
\sum_{w \in \{0,1\}^{m-n}} \|\alpha_{xw}\|^2.
$$

The quantity $q_x(|\phi_t^c\ra)$ can be viewed as the amount of amplitude 
which the network $\N$ invests in the query string $x$ to $QMQ_c$ at time 
$t.$  Intuitively, the final outcome of $\N$'s computation cannot depend 
very much on the oracle's responses to queries which have little amplitude 
invested in them.  Bennett et al. formalized this intuition 
in the following theorem (\cite{BBB+97}, Theorem 3.3):

\begin{theorem} \label{thm:3.3}
Let $|\phi_t^c \ra$ be defined as above.  Let $F \subseteq \{0,\dots,
T-1\} \times \{0,1\}^n$ be a set of time-string pairs such that
$\sum_{(t,x) \in F} q_x(|\phi_t^c \ra) \leq {\frac {\epsilon^2} T}.$
Now suppose the answer to each query instance $(t,x) \in F$ is modified to 
some arbitrary fixed 
bit $a_{t,x}$ (these answers need not be consistent with any oracle).
Let $|\tilde{\phi}_t^c \ra$ be the state of the quantum register at
time $t$ if the oracle responses are modified as stated above.
Then $| |\phi_T^c \ra - |\tilde{\phi}_T^c \ra| \leq \epsilon.
$ 
\end{theorem}

The following lemma, which is a generalization of Corollary 3.4 from
\cite{BBB+97}, shows that no quantum learning algorithm which makes
few QMQ queries can effectively distinguish many concepts in
$C^\prime$ from the typical concept $\hat{c}.$

\begin{lemma}  \label{lem:smq}
Fix any quantum network architecture $\N$ which has query complexity $T.$
For all $\epsilon > 0$ there is a set $S \subseteq C^\prime$
of cardinality at most $T^2 |C^\prime| \hat{\gamma}^C / \epsilon^2$
such that for all $c \in C^\prime \setminus S,$
we have  
$||\phi_T^{\hat{c}} \ra - |\phi_T^{c} \ra| \leq \epsilon.$
\end{lemma}
\begin{proof}
Since $||\phi^{\hat{c}}_t\ra| = 1$ for all $t = 0,1,\dots,T-1,$ 
we have 
$\sum_{t=0}^{T-1} \sum_{x\in \{0,1\}^n} q_x(|\phi^{\hat{c}}_t \ra) = T.$ 
Let $q(|\phi^{\hat{c}}_t \ra) \in \Re^{2^n}$ be the $2^n$-dimensional 
vector which has entries indexed by strings $x \in \{0,1\}^n$ and 
which has $q_x(|\phi^{\hat{c}}_t \ra)$ as its $x$-th entry.
Note that the $L_1$ norm
$\|q(|\phi^{\hat{c}}_t \ra)\|_1$ is $1$ for all $t = 0,\dots,T-1.$
For any $c_i \in C^\prime$ let 
$q_{c_i}(|\phi^{\hat{c}}_t \ra)$ be defined as 
$\sum_{x: c_i(x) \neq \hat{c}(x)} q_x(|\phi^{\hat{c}}_t \ra).$ 
The quantity $q_{c_i}(|\phi^{\hat{c}}_t \ra)$ can be viewed as the
total query magnitude with respect to $\hat{c}$ at time $t$ of those strings 
which distinguish $c_i$ from $\hat{c}.$ 
Note that $D q(|\phi^{\hat{c}}_t\ra) \in \Re^{|C^\prime|}$ 
is an $|C^\prime|$-dimensional vector whose $i$-th element is precisely
$\sum_{x: c_i(x) \neq \hat{c}(x)} q_x(|\phi^{\hat{c}}_t \ra) = 
q_{c_i}(|\phi^{\hat{c}}_t\ra).$ Since $\|D\|_1 \leq |C^\prime| 
\cdot \hat{\gamma}^C$ and $\|q(|\phi^{\hat{c}}_t \ra)\|_1 = 1,$
by the basic property of matrix norms 
we have that $\|D q(| \phi^{\hat{c}}_t \ra) \|_1 \leq |C^\prime| 
\cdot \hat{\gamma}^C,$ i.e.
$
\sum_{c_i\in C^\prime} q_{c_i}(|\phi^{\hat{c}}_t\ra) \leq |C^\prime| 
\cdot \hat{\gamma}^C.
$
Hence
$$
\sum_{t = 0}^{T-1}
\sum_{c_i \in C^\prime} q_{c_i}(|\phi^{\hat{c}}_t\ra) \leq T |C^\prime| 
\cdot \hat{\gamma}^C.
$$
If we let 
$S = \{c_i \in C^\prime : \sum_{t=0}^{T-1} q_{c_i}(|\phi^{\hat{c}}_t \ra) 
\geq {\frac {\epsilon^2} {T}} \},$
by Markov's inequality we have
$|S| \leq T^2 |C^\prime| \hat{\gamma}^C / \epsilon^2.$  Finally, 
if $c \notin S$ then $\sum_{t=0}^{T-1} q_c(|\phi^{\hat{c}}_t\ra) \leq 
{\frac {\epsilon^2}{T}}.$  Theorem \ref{thm:3.3} then implies
that $||\phi_T^{\hat{c}} \ra - |\phi_T^{c} \ra| \leq \epsilon.$
\end{proof}

Now we can prove our quantum version of Lemma \ref{lem:gamma}.

\begin{theorem} \label{thm:qgamma}
Any quantum exact learning algorithm for $C$ must have sample complexity 
$\Omega\!\left(\left({\frac 1 {\hat{\gamma}^C}}\right)^{1/2}\right).$
\end{theorem}
\begin{proof} Suppose that $\N$ is a quantum exact
learning algorithm for $\C$ which makes at most
$T = {\frac 1 {64}} \cdot \left({\frac 1 {\hat{\gamma}^C}}\right)^{1/2}$
quantum membership queries.  If we take $\epsilon = {\frac 1 {32}},$ then
Lemma \ref{lem:smq} implies that there is a set $S \subset C^\prime$
of cardinality at most  ${\frac {|C^\prime|} 4}$ such that 
for all $c \in C^\prime \setminus S$ we have
$||\phi_T^c \ra - |\phi_T^{\hat{c}} \ra| \leq {\frac 1 {32}}.$
Let $c_1, c_2$ be any two concepts in $C^\prime \setminus S.$  
By Fact \ref{fact:3.1}, the probability that $\N$
outputs a circuit equivalent to $c_1$ 
can differ by at most ${\frac 1 8}$ if $\N$'s oracle gates are 
$QMQ_{\hat{c}}$ as opposed to $QMQ_{c_1},$ and likewise for
$QMQ_{\hat{c}}$ versus $QMQ_{c_2}.$ It follows that
the probability that $\N$ outputs a circuit equivalent to $c_1$ can differ
by at most ${\frac 1 4}$ if $\N$'s oracle gates are $QMQ_{c_1}$
as opposed to $QMQ_{c_2},$ but this contradicts the assumption that
$\N$ is a quantum exact learning algorithm for $C.$
\end{proof}

\subsubsection{A Lower Bound Based on Concept Class Size}
\label{sec:ccs}

A second reason why a concept class can require many membership queries is
its size.  Angluin \cite{Angluin88} has given the following 
lower bound, incomparable to the bound of Lemma \ref{lem:gamma},
on the number of membership queries required for classical exact learning 
(the proof is given in Appendix \ref{sec:pfs}):
\begin{lemma} \label{lem:logc}
Any (classical)
exact learning algorithm for $C$ must have sample complexity 
$\Omega(\log |C|).$
\end{lemma}

In this section we prove a variant of this lemma for the quantum model.
Our proof uses ideas from \cite{BBC+98} so we introduce some of their
notation.  Let $N = 2^n.$ For each concept $c \in C,$ let $X^c = 
(X^c_0,\dots,X^c_{N-1}) \in \{0,1\}^N$ be a vector which represents $c$
as an $N$-tuple, i.e.
$X^c_i = c(x^i)$ where $x^i \in \{0,1\}^n$ is the binary
representation of $i.$  
From this perspective we may 
identify $C$ with a subset of $\{0,1\}^N,$
and we may view a $QMQ_c$ gate as a black-box oracle for $X^c$
which maps basis state $|x^i,b,y\ra$ to $|x^i,b \p X^c_i, y\ra.$

Using ideas from \cite{FFK+93,FortnowRogers98},
Beals et al. have proved the following useful 
lemma, which relates the query complexity of a quantum network
to the degree of a certain polynomial
(\cite{BBC+98}, Lemma 4.2):
\begin{lemma}  \label{lem:BBClem}
Let $\N$ be a quantum network that makes $T$ queries to 
a black-box $X,$ and let $B \subseteq \{0,1\}^m$ be a set of basis
states.  Then there exists a real-valued multilinear polynomial
$P_B(X)$ of degree at most $2T$ which equals the probability that
observing the final state of the network with black-box $X$
yields a state from $B.$
\end{lemma}

We use Lemma \ref{lem:BBClem} to prove the following quantum lower bound based
on concept class size:

\begin{theorem} \label{thm:qlogc}
Any exact quantum learning algorithm for $C$ must have sample 
complexity 
$\Omega\!\left({\frac {\log |C|} {n}}\right).$ 
\end{theorem}
\begin{proof}
Let ${\N}$ be a quantum network which learns $C$ and has query
complexity $T.$  For all $c \in C$ we have the following:
if $\N$'s oracle gates are $QMQ_c$ gates, then with probability
at least $2/3$ the output of $\N$ is a representation of a
Boolean circuit $h$ which computes $c.$ 
Let $c_1,\dots,c_{|C|}$ be all of the concepts in $C,$
and let $X^1,\dots,X^{|C|}$ be the corresponding vectors in $\{0,1\}^N.$
For all $i = 1,\dots,|C|$ let
$B_i \subseteq \{0,1\}^m$ be the collection of those basis states
which are such that if the final observation performed
by $\N$ yields a state from $B_i,$ then the output of $\N$
is a representation of a Boolean circuit which computes $c_i.$
Clearly for $i \neq j$ the sets $B_i$ and $B_j$ are disjoint.
By Lemma \ref{lem:BBClem}, for each $i = 1,\dots,|C|$ there is a real-valued
multilinear polynomial $P_i$ of degree at most $2T$ 
such that for all $j = 1,\dots,|C|,$ the value of $P_i(X^{j})$
is precisely the probability that the final observation on $\N$
yields a representation of a circuit which computes $c_i,$
provided that the oracle gates are $QMQ_{c_j}$ gates.  The
polynomials $P_i$ thus have the following properties:
\begin{enumerate}
\item $P_i(X^i) \geq 2/3$ for all $i = 1,\dots,|C|$;
\item For any $j= 1,\dots,|C|,$ we have
$ \sum_{{i \neq j}} P_i(X^{j}) \leq 1/3$
(since the total probability across all possible observations is
1).
\end{enumerate}

Let $N_0 = \sum_{i=0}^{2T} {N \choose i}.$ For any
$X = (X_0,\dots,X_{N-1})\in \{0,1\}^N$ let $\ol{X} \in \{0,1\}^{N_0}$
be the column vector which has a coordinate for each monic multilinear 
monomial over $X_0,\dots,X_{N-1}$ of degree at most $2T.$
Thus, for example, if $N=4$ and $2T = 2$
we have $X = (X_0,X_1,X_2,X_3)$ and
$$
\ol{X}^t = (1,X_0,X_1,X_2,X_3,X_0X_1,X_0X_2,X_0X_3,X_1X_2,X_1X_3,X_2X_3).
$$
If $V$ is a column vector in $\Re^{N_0},$ then $V^t \tilde{X}$ 
corresponds to the degree-$2T$ polynomial whose coefficients are given 
by the entries of $V.$  
For $i = 1,\dots,|C|$ let $V_i \in \Re^{N_0}$ be the column vector
which corresponds to the coefficients of the polynomial $P_i.$
Let $M$ be the $|C| \times N_0$ matrix whose $i$-th row is 
$V_i^t;$ note that
multiplication by $M$ defines a linear transformation from 
$\Re^{N_0}$ to $\Re^{|C|}$. 
Since $V_i^t \tilde{X}^j$ is precisely $P_i(X^j),$ the
product $M \tilde{X}^j$ is a column vector in $\Re^{|C|}$ 
which has $P_i(X^j)$ as its $i$-th coordinate.

Now let $L$ be the $|C| \times |C|$ matrix whose $j$-th column is the vector
$M \tilde{X}^j.$  
A square matrix $A$ is said to be {\em diagonally dominant}
if $|a_{ii}| > \sum_{j \neq i} |a_{ij}|$ for all $i.$ 
Properties (1) and (2) above imply that the 
transpose of $L$ is diagonally dominant.  It is well known that
any diagonally dominant matrix must be of full rank
(a proof is given in Appendix \ref{sec:dd}).
Since $L$ is full rank and each column of $L$ is in the
image of $M,$ it follows that
the image under $M$ of $\Re^{N_0}$ is all of $\Re^{|C|},$
and hence $N_0 \geq |C|.$
Finally, since $N_0 = \sum_{i=0}^{2T} {N \choose i} \leq N^{2T},$ we have
$T \geq {\frac {\log |C|} {2 \log N}} = {\frac {\log |C|} {2n}},$
which proves the theorem.
\end{proof}

The lower bound of Theorem \ref{thm:qlogc} 
is nearly tight as witnessed by the following example: let
$C$ be the collection of all $2^n$ parity functions over $\{0,1\}^n,$ so each
function in $C$ is defined by a string $a \in \{0,1\}^n$ and $c_a(x) = 
a \cdot x.$  The quantum algorithm which solves the well-known 
Deutsch-Jozsa problem \cite{DeutschJozsa92} can be used to 
exactly identify $a$ and thus learn the target concept
with probability 1 from a single query.  It follows that the factor of
$n$ in the denominator of Theorem \ref{thm:qlogc} cannot be replaced by
any function $g(n) = o(n).$

\subsection{Quantum and Classical Exact Learning are Equivalent}
\label{sec:qcele}

We have seen two different reasons why exact learning
a concept class can require a large number of (classical) membership
queries:  the class may contain many similar concepts (i.e. $\hat{\gamma}^C$ 
is small), or the class may contain very many concepts 
(i.e. $\log |C|$ is large).  The following lemma, which is a variant of
Theorem 3.1 from \cite{Hegedus95}, 
shows that these are the only reasons why
many membership queries may be required (the proof is given in
Appendix \ref{sec:pfs}).

\begin{lemma} \label{lem:ubboth}
There is an exact learning algorithm for $C$ which has sample complexity
$O((\log |C|)/ \hat{\gamma}^{C}).$
\end{lemma}

Using this upper bound we can prove that up to polynomial factors, 
quantum exact learning
is no more powerful than classical exact learning.

\begin{theorem} \label{thm:mainexact}
Let $\C$ be any concept class.  If $\C$ is exact learnable
from quantum membership queries, then $\C$ is exact
learnable from classical membership queries.
\end{theorem}
\begin{proof}
Suppose that ${\cal C}$ is not exact learnable from classical 
membership queries, i.e. for any polynomial $p$ there are infinitely
many values of $n$ such that any learning algorithm for $C_n$
requires more than $p(n)$ queries in the worst case.
By Lemma \ref{lem:ubboth}, this means that
for any polynomial $p$ there are infinitely many values of 
$n$ such that $(\log |C_n|)/\hat{\gamma}^{C_n} > p(n).$  At least
one of the following conditions must hold:
(1) for any polynomial $p$ there are infinitely many
values of $n$ such that $p(n) < 1/ \hat{\gamma}^{C_n};$ or 
(2) for any polynomial $p$ there are infinitely many
values of $n$ such that $p(n) < \log |C_n|.$  
Theorems \ref{thm:qgamma} and \ref{thm:qlogc} show that in either case
$\C$ cannot be exact learnable from a polynomial number of quantum
membership queries.
\end{proof}

In the opposite direction, 
it is easy to see that a $QMQ_c$ oracle can be used to simulate the 
corresponding $MQ_c$ oracle, so any concept class which is 
exact learnable from classical membership queries is also 
exact learnable from quantum membership queries.  This
proves Theorem \ref{thm:equivexact}.

\subsection{Discussion} \label{sec:qexactcompar}
Theorem \ref{thm:mainexact} provides an interesting
contrast to several known results for black-box quantum computation.
Let $F$ denote the set of all $2^{2^n}$ functions from $\{0,1\}^n$
to $\{0,1\}.$
Beals et al. \cite{BBC+98} have shown that if $f: F \rightarrow
\{0,1\}$ is any total function (i.e.  
$f(c)$ is defined for every possible concept $c$ over $\{0,1\}^n$), then 
the query complexity of any quantum network which computes $f$ is polynomially
related to the number of classical
black-box queries required to compute $f.$
This result is interesting because it is well known 
\cite{BernsteinVazirani97,BrassardHoyer97,DeutschJozsa92,Simon97}
that for certain concept classes $C \subset F$ and partial functions 
$f: C \rightarrow \{0,1\},$ the quantum black-box query complexity of $f$
can be exponentially smaller than the classical black-box query complexity.

Our Theorem \ref{thm:mainexact} provides a sort of dual to the results
of Beals et al.:  their bound on query complexity holds only for the fixed
concept class $F$ but for any function $f: F \rightarrow \{0,1\},$ 
while our bound holds for any concept class $C \subseteq F$ but only
for the fixed problem of exact learning.  In general, the problem of 
computing a function $f: C \rightarrow \{0,1\}$ from black-box queries
can be viewed as an ``easier'' 
version of the corresponding exact learning problem:
instead of having to figure out only one bit of information about the unknown
concept $c$ (the value of $f$), in the learning framework
the algorithm must identify $c$ exactly.
Theorem \ref{thm:mainexact} shows that for this more demanding problem,
unlike the results in 
\cite{BernsteinVazirani97,BrassardHoyer97,DeutschJozsa92,Simon97}
there is no ``clever'' way of restricting the concept class $C$ 
so that learning becomes substantially easier in the
quantum setting than in the classical setting.

\section{PAC Learning from a Quantum Example Oracle} \label{sec:qpac}

\subsection{The Quantum Example Oracle} \label{sec:qex}

Bshouty and Jackson \cite{BshoutyJackson99} have introduced
a natural quantum generalization of the standard PAC-model example
oracle.  While a standard PAC example oracle $EX(c,{\D})$ generates each 
example $\la x, c(x) \ra$ with probability
${\cal D}(x),$ where $\D$ is a distribution over $\{0,1\}^n,$
a {\em quantum PAC example oracle} $QEX(c,{\cal D})$
generates a superposition of all labeled examples,
where each labeled example $\la x, c(x) \ra$
appears in the superposition with 
amplitude proportional to the square root of ${\cal D}(x).$  
More formally, a $QEX(c,\D)$ gate
maps the initial basis state $|0^n,0\ra$ to the state $\sum_{x \in \{0,1\}^n} 
\sqrt{\D(x)} |x,c(x)\ra.$  (We leave the action of a $QEX(c,\D)$ gate
undefined on other basis states, and stipulate that 
any quantum network
which includes $T$ $QEX(c,\D)$ gates must have all $T$ gates at the ``bottom
of the circuit,'' i.e. no gate may occur on any wire between 
the inputs and any $QEX(c,\D)$ gate.)
A quantum network with $T$ $QEX(c,\D)$ gates is said to be a 
QEX network with {\em query complexity} 
$T.$

A {\em quantum PAC learning algorithm for $\C$} is a family 
$\{\N_{(n,\epsilon,\delta)} : \ n \geq 1, \  0 < \epsilon,\delta < 1\}$
of QEX networks with the following property:
for all $n \geq 1$ and $0 < \epsilon,\delta < 1,$ 
for all $c \in C_n,$ for all distributions $\D$ over $\{0,1\}^n,$ if 
the network $\N_{(n,\epsilon,\delta)}$ has all its oracle gates instantiated 
as $QEX(c,\D)$ gates, then with probability at least $1 - \delta$
the network $\N_{(n,\epsilon,\delta)}$ outputs a representation
of a circuit $h$ which is an $\epsilon$-approximator
to $c$ under ${\cal D}.$  The {\em quantum sample complexity} 
$T(n,\epsilon,\delta)$ of a quantum PAC algorithm is
the query complexity of $\N_{(n,\epsilon,\delta)}.$ A concept class
$\C$ is {\em quantum PAC learnable}
if there is a quantum PAC learning algorithm for $\C$ which has
poly$(n,{\frac 1 \epsilon},{\frac 1 \delta})$ sample complexity, 
and we say that 
$\C$ is {\em efficiently quantum PAC learnable}
if each network $N_{(n,\epsilon,\delta)}$ is of size poly$(n,
{\frac 1 \epsilon},{\frac 1 \delta}).$  

\subsection{Lower Bounds on Classical and Quantum PAC Learning}

Throughout this section 
for ease of notation we omit the subscript $n$ and write $C$ for $C_n.$
We view each concept $c \in C$ as a subset of $\{0,1\}^n.$
For $S \subseteq \{0,1\}^n,$ we write $\Pi_{C}(S)$ to denote
$\{c \cap S : c \in C\},$ so $|\Pi_C(S)|$ is the number of different 
``dichotomies'' which the concepts in $C$ induce 
on the points in $S.$ A subset 
$S \subseteq \{0,1\}^n$ is said to be {\em shattered} by $C$ 
if $|\Pi_C(S)| = 2^{|S|},$ i.e. if $C$ induces
every possible dichotomy on the points in $S.$  
The {\em Vapnik-Chervonenkis dimension of $C$,} VC-DIM$(C),$ is the 
size of the largest subset $S \subseteq \{0,1\}^n$ which is 
shattered by $C.$

Well-known results in computational learning theory show that the
Vapnik-Chervonenkis dimension of a concept class $C$ characterizes the
number of calls to $EX(c,\D)$ which are information-theoretically
necessary and sufficient to PAC learn $C.$
For the lower bound, the following theorem is (a slight
simplification of) a result due to Blumer et al. (\cite{BEHW89},
Theorem 2.1.ii.b); a proof sketch is given in Appendix \ref{sec:pfs}.
(A stronger bound was later given by Ehrenfeucht et al. \cite{EHKV89}.) 

\begin{theorem} \label{thm:behw}
Let $C$ be any concept class and $d =$ VC-DIM$(C).$
Then any (classical) 
PAC learning algorithm for $\C$ must have sample complexity $\Omega(d).$
\end{theorem} 

The following theorem is a quantum analogue of Theorem \ref{thm:behw};
the proof, which extends the techniques used in the 
proof of Theorem \ref{thm:qlogc} using ideas from error-correcting codes,
is given in Appendix 
\ref{sec:vcq}.

\begin{theorem} \label{thm:behwq}
Let $\C$ be any concept class and $d =$ VC-DIM$(C).$
Then any quantum PAC learning algorithm for $\C$ must have quantum
sample complexity $\Omega({\frac d n}).$
\end{theorem} 

Since the class of parity functions over $\{0,1\}^n$ has Vapnik-Chervonenkis
dimension $n,$ as in Section \ref{sec:ccs} 
the factor of $n$ in the denominator
of Theorem \ref{thm:behwq} cannot be replaced by any function $g(n) = o(n).$

\subsection{Quantum and Classical PAC Learning are Equivalent}

A well-known theorem due to 
Blumer et al. (Theorem 3.2.1.ii.a of \cite{BEHW89}) 
shows that the VC-dimension of a concept
class bounds the number of $EX(c,\D)$ calls 
required for (classical) PAC learning:
\begin{theorem} \label{thm:behwmain}
Let $C$ be any concept class and $d =$ VC-DIM$(C).$  
There is a (classical) PAC learning algorithm for $C$ which has sample
complexity $O({\frac 1 \epsilon} \log {\frac 1 \delta} + 
{\frac d \epsilon} \log {\frac 1 \epsilon}).$
\end{theorem}

The proof of Theorem \ref{thm:behwmain} is quite complex so we do
not attempt to sketch it.
As in Section \ref{sec:qcele}, this upper bound along with 
our lower bound from Theorem \ref{thm:behwq}
together yield:

\begin{theorem} \label{thm:mainpac}
Let $\C$ be any concept class.  If $\C$ is quantum PAC learnable, 
then $\C$ is (classically) PAC learnable.
\end{theorem}

A $QEX(c,\D)$ oracle can be used to simulate the corresponding $EX(c,\D)$
oracle by immediately performing an observation on the $QEX$ gate's outputs;
such an observation yields each example $\la x,c(x) \ra$ with probability
$\D(x)$.\footnote{ 
As noted in Section \ref{sec:qc}, intermediate
observations during a computation can always be simulated by a single
observation at the end of the computation.}  Consequently
any concept class which is classically 
PAC learnable is also quantum PAC learnable,
and Theorem \ref{thm:equivpac} is proved.

\section{Quantum versus Classical Efficient Learnability} \label{sec:qeff}
We have shown that from an information-theoretic perspective, quantum
learning is no more powerful than classical learning (up to polynomial
factors).  However, we now observe that the apparant {\em computational}
advantages of the quantum model yield efficient quantum learning
algorithms which are believed to have no efficient classical counterparts.

A {\em Blum integer} is an integer $N = pq$ where $p \neq q$ are
$\ell$-bit primes each congruent to 3 modulo 4.
It is widely believed that there is no polynomial-time
classical algorithm which can successfully factor a randomly
selected Blum integer with nonnegligible success probability.

Kearns and Valiant \cite{KearnsValiant94} have constructed
a concept class $\C$ with the following property:
a polynomial-time (classical)
PAC learning algorithm for $\C$ would yield a
polynomial-time algorithm for factoring Blum integers.  
Thus, assuming that factoring Blum integers is a 
computationally hard problem for classical computation,
the Kearns-Valiant concept class $\C$ is not efficiently PAC learnable.
On the other hand, in a celebrated result Shor \cite{Shor97}
has exhibited a poly$(n)$ size quantum network which can factor
an arbitrary $n$-bit integer with high success probability.  His
construction yields an efficient quantum PAC learning
algorithm for the Kearns-Valiant concept class.  We thus have
\begin{observation} \label{obs:o1}
If there is no polynomial-time classical algorithm for factoring 
Blum integers, then there is a concept class $\C$ which is efficiently
quantum PAC learnable but not efficiently classically PAC learnable.
\end{observation}

The hardness results of Kearns and Valiant were later extended by 
Angluin and Kharitonov \cite{AngluinKharitonov95}.  Using a 
public-key encryption system which is secure against chosen-cyphertext
attack (based on the assumption that factoring Blum integers is 
computationally hard for polynomial-time algorithms), they constructed
a concept class $\C$ which cannot be learned by any polynomial-time
learning algorithm which makes membership queries.  As with
the Kearns-Valiant concept class, though, using Shor's quantum
factoring algorithm it is possible to construct an efficient
quantum exact learning algorithm for this concept class.
Thus, for the exact learning model as well, we have:
\begin{observation} \label{obs:o2}
If there is no polynomial-time classical algorithm for factoring 
Blum integers, then there is a concept class $\C$ which is efficiently
quantum exact learnable from membership queries but not efficiently 
classically exact learnable from membership queries.
\end{observation}

\section{Conclusion and Future Directions}
While we have shown that quantum and classical learning are (up to
polynomial factors) information-theoretically
equivalent, many interesting questions remain about
the relationship between efficient quantum and classical
learnability.  One goal is to
prove analogues of Observations \ref{obs:o1} and \ref{obs:o2}
under a weaker computational hardness assumption such as the existence
of any one-way function; it seems plausible that some
some combination of cryptographic techniques together with the ideas used
in Simon's quantum algorithm \cite{Simon97} might be able
to achieve this.  Another goal is to develop efficient quantum
learning algorithms for natural concept classes, such as the polynomial-time
quantum algorithm of Bshouty and Jackson \cite{BshoutyJackson99} 
for learning DNF formulae from uniform quantum examples.

\appendix

\section{Bounds on Classical Sample Complexity} \label{sec:pfs}

\noindent {\bf Proof of Lemma \ref{lem:gamma}}:
Let $C^\prime \subseteq C,$ $|C^\prime| \geq 2$ be such that
$\gamma^{C^\prime} = 
\hat{\gamma}^C.$
Consider the following adversarial
strategy for answering queries: given the query string $a,$
answer the bit $b$ which maximizes $\gamma^{C^\prime}_{\la a,b \ra}.$
This strategy ensures that each response eliminates
at most a $\gamma^{C^\prime}_a \leq \gamma^{C^\prime} = 
\hat{\gamma}^{C}$ fraction of the concepts in $C^\prime.$
After ${\frac 1 {2 \hat{\gamma}^C}} - 1$ membership
queries, fewer than half of the concepts in $C^\prime$ have been 
eliminated, so at least two concepts have not yet been eliminated.  
Consequently, it is impossible for $A$ to output a hypothesis which is 
equivalent to the correct concept with probability greater than $1/2.$ 
\qqed{(Lemma \ref{lem:gamma})~}

\bigskip

\noindent {\bf Proof of Lemma \ref{lem:logc}}:  
Consider the following adversarial strategy for answering queries:
if $C^\prime \subseteq C$ is the set of concepts 
which have not yet been
eliminated by previous responses to queries, then given the query
string $a,$ answer the bit $b$ such that 
$\gamma^{C^\prime}_{\la a,b \ra} \geq {\frac 1 2}.$
Under this strategy, after $\log |C| - 1$ membership queries
at least two possible target concepts will remain.
\qqed{(Lemma \ref{lem:logc})~}

\bigskip

\noindent {\bf Proof of Lemma \ref{lem:ubboth}}:  
Consider the following (classical) learning algorithm $A$:  at each stage in 
its execution, if $C^\prime$ is the set of 
concepts in $C$ which have not yet been eliminated by previous responses
to queries, algorithm $A$'s next query string is the string
$a \in \{0,1\}^n$ which maximizes $\gamma^{C^\prime}_a.$  
By following this strategy, each query response received from the oracle
must eliminates at least a  $\gamma^{C^\prime}$
fraction of the set $C^\prime,$ so 
with each query the size of the set of possible target concepts
is multiplied by a factor which is at most
$1 - \gamma^{C^\prime} \leq 1 - \hat{\gamma}^C.$  Consequently,
after $O((\log |C|)/\hat{\gamma}^C)$ queries, only a single concept will
not have been eliminated; this concept must be the target concept,
so $A$ can output a hypothesis $h$ which is equivalent to $c.$
\qqed{(Lemma \ref{lem:ubboth})~}

\bigskip

\noindent {\bf Proof Sketch for Theorem \ref{thm:behw}:}
The idea behind Theorem \ref{thm:behw}
is to consider the distribution $\D$ which is uniform over
some shattered set $S$ of size $d$ and assigns zero weight to points
outside of $S.$  Any learning algorithm which makes only
$d/2$ calls to $EX(c,\D)$ will have no information about 
the value of $c$ on at least $d/2$ points in $S;$ moreover, since
the set $S$ is shattered by $C,$ any labeling is possible 
for these unseen points.  Since the error of
any hypothesis $h$ under $\D$ is the fraction of points in $S$ 
where $h$ and the target concept disagree, 
a simple analysis shows that no learning algorithm which perform only $d/2$ 
calls to $EX(c,\D)$ can have high probability (e.g. 
$1 - \delta = 2/3$) of generating a low-error hypothesis 
(e.g. $\epsilon = 1/10$).
\qqed{(Theorem \ref{thm:behw})~}

\section{Proof of Theorem \ref{thm:behwq}} \label{sec:vcq}

Let $S = \{x^1,\dots,x^d\}$ be a set which is shattered by $C$ and let 
$\D$ be the distribution which is uniform on $S$ and assigns zero
weight to points outside $S.$
If $h: \{0,1\}^n \rightarrow \{0,1\}$ is a Boolean function on $\{0,1\}^n,$ 
we say that the {\em relative distance of $h$ and $c$ on $S$} is the
fraction of points in $S$ on which $h$ and $c$ disagree.
We will prove the following result
which is stronger than Theorem \ref{thm:behwq}:  Let $\N$ be a 
quantum network with $QMQ$ gates such that for all $c \in C,$ 
if $\N$'s oracle gates are $QMQ_c$ gates, then with probability at least 
$2/3$ the output of $\N$ is a hypothesis $h$ 
such that the relative distance of $h$ and $c$ on $S$ is at most $1/10.$ 
We will show that such a network $\N$ must have query complexity at least 
${\frac d {12 n}}.$
Since any QEX network with query complexity $T$ can
be simulated by a QMQ network with
query complexity $T,$ taking $\epsilon = 1/10$ and $\delta = 1/3$
will prove Theorem \ref{thm:behwq}.

The argument is a modification of the proof of Theorem \ref{thm:qlogc}.
Let $\N$ be a quantum network with query complexity $T$
which satisfies the following condition:  for all 
$c \in C,$ if $\N$'s oracle gates are $QMQ_c$ gates, then with
probability at least $2/3$ the output of $\N$ is a representation
of a Boolean circuit $h$ such that the relative distance of $h$
and $c$ on $S$ is at most $1/10.$  By the well-known
Gilbert-Varshamov bound from coding theory
(see, e.g., Theorem 5.1.7 of \cite{VanLint92}),
there exists a set $s^1,\dots,s^A$ of $d$-bit
strings such that for all $i \neq j$ the strings $s^i$ and $s^j$
differ in at least $d/4$ bit positions, where 
$$
A \geq {\frac {2^d} {\sum_{i = 0}^{d/4 - 1} {d \choose i}}}
\geq {\frac {2^d} {\sum_{i = 0}^{d/4} {d \choose i}}}
\geq 2^{d(1 - H(1/4))} > 2^{d/6}.
$$
(Here $H(p) = -p \log p - (1-p) \log(1-p)$ is the binary entropy
function.)  For each $i = 1,\dots,A$ let $c_i \in C$ be a
concept such that the $d$-bit string $c_i(x^1)\cdots c_i(x^d)$
is $s^i$ (such a concept $c_i$ must exist since the set
$S$ is shattered by $C$).

For $i=1,\dots,A$ let $B_i \subseteq \{0,1\}^m$
be the collection of those basis states which are such that
if the final observation performed by $\N$ yields a state from $B_i,$
then the output of $\N$ is a hypothesis $h$ such that
$h$ and $c_i$ have relative distance at most $1/10$ on $S.$  Since
each pair of concepts $c_i,c_j$ has relative distance at least
$1/4$ on $S,$ the sets $B_i$ and $B_j$ are disjoint for all $i \neq j.$

As in Section \ref{sec:lbcqel} let $N = 2^n$ and let
$X^j = (X^j_0,\dots,X^j_{N-1})
\in \{0,1\}^n$ where $X^j$ is the $N$-tuple representation
of the concept $c_j.$
By Lemma \ref{lem:BBClem}, for each $i = 1,\dots,A$ there is
a real-valued multilinear polynomial $P_i$ of degree at most
$2T$ such that for all $j=1,\dots,A,$ the value of $P_i(X^j)$
is precisely the probability that the final observation on $\N$
yields a state from $B_i$
provided that the oracle gates are $QMQ_{c_j}$ gates.
Since, by assumption, if $c_i$ is the target concept then
with probability at least $2/3$ 
$\N$ generates a hypothesis which has relative distance at most $1/10$
from $c_i$ on $S,$ the polynomials $P_i$ have the following properties:
\begin{enumerate}
\item $P_i(X^i) \geq 2/3$ for all $i = 1,\dots,A;$
\item For any $j = 1,\dots,A$ we have that $\sum_{i \neq j} P_i(X^j)
\leq 1/3$ (since the $B_i$'s are disjoint and the total probability
across all observations is 1).
\end{enumerate}

Let $N_0$ and $\tilde{X}$ be defined as in the proof of 
Theorem \ref{thm:qlogc}.  For $i = 1,\dots,A$ let $V_i \in \Re^{N_0}$
be the column vector which corresponds to the coefficients of the polynomial
$P_i,$ so $V_i^t \tilde{X} = P_i(X).$  Let $M$
be the $A \times N_0$ matrix whose $i$-th row is
the vector $V_i^t,$ so multiplication by $M$ is
a linear transformation from $\Re^{N_0}$ to $\Re^A.$
The product $M \tilde{X}^j$ is a column vector in
$\Re^{A}$ which has $P_i(X)$ as its $i$-th coordinate.

Now let $L$ be the $A \times A$ matrix whose $j$-th column is the vector
$M \tilde{X}^j.$  As in Theorem \ref{thm:qlogc} we have that
the transpose of $L$ is diagonally dominant, so $L$ is of
full rank and hence $N_0 \geq A.$  Since $A \geq 2^{d/6}$
we thus have that 
$T \geq {\frac {d/6} {2 \log_2 N}} = {\frac {d} {12n}},$
and the theorem is proved.
\qqed{(Theorem \ref{thm:behwq})~}

\section{A diagonally dominant matrix has full rank} \label{sec:dd}

This fact follows from the following theorem (see, e.g.,
Theorem 6.1.17 of \cite{Ortega87}).

\begin{theorem} [Gershgorin's Circle Theorem] 
\label{thm:gersh}
Let $A$ be a real or complex-valued $n \times n$ matrix.  
Let $S_i$ be the disk in the complex plane whose center is $a_{ii}$
and whose radius is $r_i = \sum_{j \neq i} |a_{ij}|.$  Then every
eigenvalue of $A$ lies in the union of the
disks $S_1,\dots,S_n.$
\end{theorem}
\begin{proof}
If $\lambda$ is an eigenvalue of $A$ which has corresponding
eigenvector $x = (x_1,\dots,x_n),$ then since $Ax = \lambda x$
we have 
$$
(\lambda - a_{ii})x_i = \sum_{j \neq i} a_{ij} x_j \mbox{~~for~}i=1,\dots,n.
$$
Without loss of generality we may assume that 
$\|x\|_{\infty} = 1,$ so $|x_k| = 1$ for some $k$
and $|x_j| \leq 1$ for $j \neq k.$  Thus 
$$
|\lambda - a_{kk}| = |(\lambda - a_{kk})x_k| \leq \sum_{j \neq k}
|a_{kj}||x_j| \leq \sum_{j \neq k} |a_{kj}|
$$
and hence $\lambda$ is in the disk $S_k.$
\end{proof}

For a diagonally dominant matrix the radius $r_i$ of each disk $S_i$ 
is less than its distance from the origin, which is $|a_{ii}|.$  Hence
$0$ cannot be an eigenvalue of a diagonally dominant matrix, so
the matrix must have full rank.


\begin{thebibliography}{99}
\begin{small}
\bibitem{Angluin88} D. Angluin.  Queries and concept learning,
{\em Machine Learning} {\bf 2} (1988), 319-342.

\bibitem{AngluinKharitonov95} 
D. Angluin and M. Kharitonov.  When won't membership
queries help?  {\em J. Comp. Syst. Sci.} {\bf 50} (1995), 336-355.

\bibitem{AnthonyBiggs97} M. Anthony and N. Biggs.  {\em Computational
Learning Theory:  an Introduction.} Cambridge Univ. Press, 1997.

\bibitem{BBC+98} R. Beals, H. Buhrman, R. Cleve, M. Mosca and R. de Wolf.
Quantum lower bounds by polynomials, {\em in} ``Proc. 39th IEEE Symp.
on Found. of Comp. Sci.,'' (1998), 352-361. quant-ph/9802049.

\bibitem{BBB+97} C. Bennett, E. Bernstein, G. Brassard and U.
Vazirani.  Strengths and weaknesses of quantum computing, {\em
SIAM J. Comput.} {\bf 26}(5) (1997), 1510-1523.

\bibitem{BernsteinVazirani97} E. Bernstein and U. Vazirani.
Quantum complexity theory, {\em SIAM J. Comput.,} {\bf 26}(5) 
(1997), 1411-1473.

\bibitem{BEHW89} A. Blumer, A. Ehrenfeucht, D. Haussler and M. K.
Warmuth.
Learnability and the Vapnik-Chervonenkis
Dimension, {\em J. ACM} {\bf 36}(4) (1989), 929-965.

\bibitem{BBH+98} M. Boyer, G. Brassard, P. H\o yer, A. Tapp.  Tight 
bounds on quantum searching,
{\em Forschritte der Physik} {\bf 46}(4-5) (1998), 493-505.  

\bibitem{BHT98} G. Brassard, P. H\o yer and A. Tapp.  Quantum counting,
{\em in} ``Proc. 25th ICALP'' (1998) 820-831.  quant-ph/9805082.

\bibitem{BrassardHoyer97} G. Brassard and  P. H\o yer.  An exact
quantum polynomial-time algorithm for Simon's problem, 
{\em in} ``Fifth Israeli Symp. on Theory of Comp. and Systems''
(1997), 12-23.

\bibitem{BCG+96} N. Bshouty, R. Cleve, R. Gavald\`{a}, S. Kannan and
C. Tamon.  Oracles and queries that are sufficient for exact
learning, {\em J. Comput. Syst. Sci.} {\bf 52}(3) (1996), 421-433.

\bibitem{BshoutyJackson99} N. Bshouty and J. Jackson. Learning
DNF over the uniform distribution using a quantum example oracle,
{\em SIAM J. Comput.} {\bf 28}(3) (1999), 1136-1153.

\bibitem{BCW98} H. Buhrman, R. Cleve and A. Wigderson.  Quantum vs. classical
communication and computation, {\em in} ``Proc. 30th ACM Symp. on
Theory of Computing,'' (1998), 63-68. quant-ph/9802040.

\bibitem{Cleve99} R. Cleve.  An introduction to quantum complexity
theory, {\em to appear in} ``Collected Papers on Quantum
Computation and Quantum Information Theory,'' ed. by C. Macchiavello, G.M.
Palma and A. Zeilinger.  quant-ph/9906111.

\bibitem{DeutschJozsa92} D. Deutsch and R. Jozsa.  Rapid
solution of problems by quantum computation, {\em Proc. Royal
Society of London A,} {\bf 439} (1992), 553-558.

\bibitem{EHKV89} A. Ehrenfeucht, D. Haussler, M. Kearns
and L. Valiant.  A general lower bound on the number of 
examples needed for learning, {\em Inf. and Comput.} {\bf 82} (1989),
246-261.

\bibitem{FFK+93} S. Fenner, L. Fortnow, S. Kurtz and L. Li. An oracle
builder's toolkit, {\em in} ``Proc. Eigth Structure in Complexity
Theory Conference'' (1993), 120-131.

\bibitem{FortnowRogers98} L. Fortnow and J. Rogers.  Complexity limitations on
quantum computation, {\em in} ``Proc. 13th Conf. on Computational
Complexity'' (1998), 202-209.

\bibitem{Gavalda94} R. Gavald\`{a}. The complexity of learning with queries,
{\em in} ``Proc. Ninth Structure in Complexity Theory Conference'' (1994),
324-337.

\bibitem{Grover96} L. K. Grover.  A fast quantum mechanical algorithm
for database search, {\em in} ``Proc. 28th Symp. on Theory of Computing''
(1996), 212-219.

\bibitem{Hegedus95} T. Heged\H{u}s.  Generalized teaching dimensions
and the query complexity of learning, {\em in}
``Proc. Eigth Conf. on Comp. Learning Theory,'' (195), 108-117.

\bibitem{HPR+96} L. Hellerstein, K. Pillaipakkamnatt, V.
Raghavan and D. Wilkins.  How many queries are needed to learn?  {\em J. ACM}
{\bf 43}(5) (1996), 840-862.

\bibitem{KearnsValiant94} M. Kearns and L. Valiant.
Cryptographic limitations on learning boolean formulae and finite
automata, {\em J. ACM} {\bf 41}(1) (1994), 67-95.

\bibitem{KearnsVazirani94} M. Kearns and U. Vazirani.  {\em An
Introduction to Computational Learning Theory.} MIT Press, 1994.

\bibitem{Ortega87} J. Ortega.  {\em Matrix Theory: a second
course.} Plenum Press, 1987.

\bibitem{Shor97} P. Shor.  Polynomial-time algorithms for
prime factorization and discrete logarithms on a
quantum computer, {\em SIAM J. Comput.} {\bf 26}(5) (1997),
1484-1509.

\bibitem{Simon97} D. Simon.  On the power of quantum computation,
{\em SIAM J. Comput.} {\bf 26}(5) (1997), 1474-1483.

\bibitem{Valiant84} L. G. Valiant. A theory of the learnable,
{\em Comm. ACM} {\bf 27}(11) (1984), 1134-1142.

\bibitem{VanLint92} J. H. Van Lint.  {\em Introduction to 
Coding Theory.} Springer-Verlag, 1992.

\bibitem{VC71} V.N. Vapnik and A.Y. Chervonenkis.  On the uniform
convergence of relative frequencies of events to their probabilities,
{\em Theory of Probability and its Applications,} {\bf 16}(2) (1971), 264-280.

\bibitem{Yao93} A.C. Yao.  Quantum circuit complexity, {\em in}
``Proc. 34th Symp. on Found. of Comp. Sci.'' (1993), 352-361.

\bibitem{Zalka97} C. Zalka.  Grover's quantum searching algorithm is
optimal.  quant-ph/9711979, Nov 1997.

\end{small}
\end{thebibliography}
\end{document}